\documentclass[10pt,conference,letterpaper]{RWWTemplate}
\IEEEoverridecommandlockouts
\renewcommand{\baselinestretch}{0.93}
\usepackage{graphicx}
\graphicspath{ {figures/} }
\def\BibTeX{{\rm B\kern-.05em{\sc i\kern-.025em b}\kern-.08em
    T\kern-.1667em\lower.7ex\hbox{E}\kern-.125emX}}
\usepackage{cite}
\usepackage{mathtools}
\usepackage{blindtext}
\usepackage{amsmath,amssymb,amsfonts}
\usepackage{algorithm}
\usepackage{algpseudocode}
\usepackage{setspace}
\usepackage{subcaption}	
\usepackage{textcomp}
\usepackage{balance}
\usepackage{xcolor}
\usepackage{multirow}
\usepackage{stfloats}
\usepackage{amsmath}
\usepackage{array}
\usepackage[hyphens]{url}
\usepackage{tabularx}
\usepackage{bm}
\usepackage{pifont}% 
\let\ssstyle=\scriptscriptstyle
\newcommand{\define}{\newcommand}
\newcommand{\bt}{\begin{tabular}}
\newcommand{\et}{\end{tabular}}
\newcommand{\be}{\begin{equation}}
\newcommand{\ee}{\end{equation}}

\newcommand{\bd}{\begin{displaymath}}
\newcommand{\ed}{\end{displaymath}}
\newcommand{\ba}{\begin{array}}
\newcommand{\ea}{\end{array}}
\newcommand{\bn}{\begin{enumerate}}
\newcommand{\en}{\end{enumerate}}
\newcommand{\bds}{\begin{description}}
\newcommand{\eds}{\end{description}}
\newcommand{\bi}{\begin{itemize}}
\newcommand{\ei}{\end{itemize}}
\newcommand{\bc}{\begin{center}}
\newcommand{\ec}{\end{center}}
\newcommand{\bqa}{\begin{eqnarray}}
\newcommand{\eqa}{\end{eqnarray}}

\define{\ul}{\noindent \underline}
\define{\Gamb}{{\bf \Gamma}}
\define{\Sigb}{{\bf \Sigma}}
\define{\Lambs}{{\bf \Lambda}_{s}}
\define{\LambR}{{\bf \Lambda}_{R}}
\define{\OmB}{{\bf \Omega}_{oB}}
\define{\Omo}{{\bf \Omega}_o}
\define{\Omb}{{\bf \Omega}}
\define{\Oms}{{\bf \Omega}_s}
\define{\alpb}{{\bf \alpha}}
\define{\alpbb}{\bar{\bf \alpha}}
\define{\betb}{{\bf \beta}}
\define{\betbb}{\bar{\bf \beta}}
\define{\tr}{\mbox{Tr}}
\define{\CR}{Cram\'{e}r-Rao~}
\define{\Mu}{{\it MUSIC}}
\define{\ES}{{\it ESPRIT}}
\define{\Mi}{{\it Min-Norm}}
\define{\SR}{{\it Subspace Rotation}}
\define{\ath}{{{\bf a}(\theta_k)}}
\define{\Ath}{{\bf A}(\theta)}
\define{\AthH}{{\bf A}^{H}(\theta)}
\define{\athd}{{{\bf a}^{(1)}(\theta_k)}}
\define{\bo}{{\bf 1}}
\define{\bz}{{\bf 0}}
\define{\ab}{{\bf a}}
\define{\bb}{{\bf b}}
\define{\cb}{{\bf c}}
\define{\cbb}{\bar{\bf c}}
\define{\db}{{\bf d}}
\define{\eb}{{\bf e}}
\define{\ebb}{\bar{\bf e}}
\define{\ebk}{{\bf e}_k}
\define{\ebbk}{\bar{\bf e}_k}
\define{\fb}{{\bf f}}
\define{\gb}{{\bf g}}
\define{\hb}{{\bf h}}
\define{\mb}{{\bf m}}
\define{\nb}{{\bf n}}
\define{\ob}{{\bf o}}
\define{\pb}{{\bf p}}
\define{\qb}{{\bf q}}
\define{\rb}{{\bf r}}
\define{\sbb}{{\bf s}}
\define{\tb}{{\bf t}}
\define{\ub}{{\bf u}}
\define{\vb}{{\bf v}}
\define{\wb}{{\bf w}}
\define{\st}{{\bf s}}
\define{\xb}{{\bf x}}
\define{\yb}{{\bf y}}
\define{\zb}{{\bf z}}
\define{\Ab}{{\bf A}}
\define{\Psib}{{\bf \Psi}}
\define{\Xib}{{\bf \Xi}}

\define\sT{{\ssstyle T}}

\define{\Au}{{\Ab^{\uparrow}}}
\define{\Af}{{\bf A}_F}
\define{\As}{{\bf A}_S}
\define{\Bb}{{\bf B}}
\define{\Cb}{{\bf C}}
\define{\Cbh}{\hat{\bf C}}
\define{\Db}{{\bf D}}
\define{\Eb}{{\bf E}}
\define{\Fb}{{\bf F}}
\define{\Gb}{{\bf G}}
\define{\Hb}{{\bf H}}
\define{\Ib}{{\bf I}}
\define{\Is}{{\bf I}_S}
\define{\Ibd}{{\bf I}^{\downarrow}}
\define{\Ibu}{{\bf I}^{\uparrow}}
\define{\Jb}{{\bf J}}
\define{\Kb}{{\bf K}}
\define{\Lb}{{\bf L}}
\define{\Lbb}{\bar{\bf L}}
\define{\Mb}{{\bf M}}
\define{\Nb}{{\bf N}}
\define{\Ob}{{\bf O}}
\define{\Pb}{{\bf P}}
\define{\Qb}{{\bf Q}}
\define{\Rb}{{\bf R}}
\define{\Rbh}{\hat{\bf R}}
\define{\Rsd}{\Delta {\bf R}_s}
\define{\Rba}{{\bf R}_A}
\define{\Rbse}{\stackrel{\sim}{{\bf R}_{s}}}
\define{\Rbb}{\bar{\bf R}}
\define{\Sb}{{\bf S}}
\define{\Tb}{{\bf T}}
\define{\Ub}{{\bf U}}
\define{\Vb}{{\bf V}}
\define{\Vn}{{\bf V_n}}
\define{\Wb}{{\bf W}}
\define{\Xb}{{\bf X}}
\define{\Yb}{{\bf Y}}
\define{\Zb}{{\bf Z}}
\define{\thb}{{\bf \Theta}}
\define{\Yd}{\Delta {\bf Y}}
\define{\Yt}{\tilde{\bf Y}}
\define{\Ut}{\tilde{\bf U}}
\define{\Vt}{\tilde{\bf V}}
\define{\Unt}{\tilde{\bf U}_o}
\define{\unt}{\tilde{\bf u}_o}
\define{\Ust}{\tilde{\bf U}_s}
\define{\ust}{\tilde{\bf u}_s}
\define{\Wt}{\tilde{\bf W}}
\define{\Wd}{\Delta \bf {W}}
\define{\Phb}{{\bf {\Phi}}}
\define{\Omegab}{{\bf {\Omega}}}
\define{\Gammab}{{\bf {\Gamma}}}
\define{\Phib}{{\bf {\Phi}}}
\define{\Phd}{\Delta \bf {\Phi}}
\define{\Phl}{{{\bf \Phi}^{\downarrow}}}
\define{\Phu}{{{\bf \Phi}^{\uparrow}}}
\define{\Psd}{\Delta \bf {\Psi}}
\define{\rt}{\tilde{r}}
\define{\rd}{\Delta r}
\define{\td}{\Delta \theta}
\define{\tl}{\tilde{\theta}}
\define{\ut}{\tilde{\bf u}}
\define{\Und}{\Delta {\bf U}_o}
\define{\Usd}{\Delta {\bf U}_s}
\define{\Us}{{\bf U}_s}
\define{\us}{{\bf u}_s}
\define{\Uo}{{\bf U}_o}
\define{\uo}{{\bf u}_o}
\define{\Vs}{{\bf V}_s}
\define{\vs}{{\bf v}_s}
\define{\Vo}{{\bf V}_o}
\define{\vo}{{\bf v}_o}
\define{\Aa}{||{\bf \alpha}_k||^{2}}
\define{\Usx}{{\bf U}_{sx}}
\define{\Usy}{{\bf U}_{sy}}
\define{\Usxd}{\Delta {\bf U}_{sx}}
\define{\Usyd}{\Delta {\bf U}_{sy}}
\define{\Usxt}{\tilde{\bf U}_{sx}}
\define{\Usyt}{\tilde{\bf U}_{sy}}
\define{\Usz}{{\bf U}_{sz}}
\define{\Uszd}{\Delta {\bf U}_{sz}}
\define{\Uszt}{\tilde{\bf U}_{sz}}
\define{\Ft}{\tilde{\bf F}}
\define{\Fd}{\Delta {\bf F}}
\define{\At}{\tilde{\bf A}}
\define{\Bt}{\tilde{\bf B}}
\define{\Bd}{\Delta {\bf B}}
\define{\ld}{\Delta \bar{\lambda}_k}
\define{\ldd}{\Delta \lambda_k}
\define{\lt}{\tilde{\lambda}}
\define{\lb}{\bar{\lambda}}
\define{\df}{\stackrel{\rm def}{=}}
\define{\diag}{\rm diag}
\define{\prf}{\noindent \underline{Proof:}\\}
\define{\pe}{\hfill $\Box$}
\define{\doi}{\stackrel{j\neq i}{j=1}}
\define{\dok}{\stackrel{j\neq k}{j=1}}
\define{\ds}{\displaystyle}
\define{\Usu}{{\bf U}_s^{\uparrow}}
\define{\Usl}{{\bf U}_s^{\downarrow}}
\define{\Unu}{{\bf U}_o^{\uparrow}}
\define{\Unl}{{\bf U}_o^{\downarrow}}
\define{\Oo}{{\bf \Omega}_{o}}
\define{\Ot}{{\tilde{\bf \Omega}}_{o}}
\define{\Os}{{\bf \Omega}_{s}}
\define{\Ou}{{\bf O}^{\uparrow}}
\define{\Od}{{\bf O}^{\downarrow}}
\define{\thd}{\Delta \theta}
\define{\Iu}{{\bf I}^{\uparrow}}
\define{\Il}{{\bf I}^{\downarrow}}
\define{\Otd}{{\tilde{\bf O}}^{\downarrow}}
\define{\Sr}{{\bf \Sigma}_s^{{1}\over{2}}}
\define{\Sp}{{\bf \Sigma}_s^{-1}}
\define{\Ss}{{\bf \Sigma}_s^{-{{1}\over{2}}}}
\define{\nun}{\underline{\bf n}}
\define{\nh}{\bar{\bf n}}
\define{\eq}[1]{(\ref{#1})}
\define{\dB}{\delta {\bf B}}
\define{\deb}{\delta b}
\define{\bul}{\underline{\beta}}
\define{\ape}{\stackrel{\cdot}{=}}

\define{\im}{\,\mbox{Im}}
\define{\re}{\,\mbox{Re}}

\define{\rmE}{\mathrm{E}}
\define{\rme}{\mathrm{e}}

\let\ssstyle=\scriptscriptstyle

\def\s0{{\ssstyle 0}}

\def\sT{{\ssstyle T}}

\define{\Pib}{{\bf \Pi}}
\define{\xz}{\epsilon_{l,0}(n)}
\define{\xx}{\epsilon_{l,1}(n)}
\define{\xc}{\epsilon_{l,2}(n)}
\newcommand\blfootnote[1]{%
		\begingroup
		\renewcommand\thefootnote{}\footnote{#1}%
		\addtocounter{footnote}{-1}%
		\endgroup
	}
\title{28~GHz mmWave Channel Measurements and Modeling in a Library Environment
\vspace{-0.3cm}
}

\author{%
Fatih Erden, Ozgur Ozdemir, and Ismail Guvenc%
\vspace{2pt}\\
Department of Electrical and Computer Engineering, NCSU, Raleigh, NC 27606\\
    \vspace{-0.8cm}
}

\begin{document}
\maketitle

\begin{abstract}
To fully exploit the millimeter-wave bands for the fifth generation cellular systems, an accurate understanding of the channel propagation characteristics is required, and hence extensive measurement campaigns in different environments are needed. In this paper, we use a rotated directional antenna-based channel sounder for measurements at 28~GHz in large indoor environments at a library setting. We present models for power angular-delay profile and large-scale path loss based on the measurements over distances ranging from 10~m to 50~m. In total, nineteen different line-of-sight~(LOS) and non-line-of-sight~(NLOS) scenarios are considered, including the cases where the transmitter and the receiver are placed on different floors. Results show that the close-in free space reference distance and the floating intercept path loss models both perform well in fitting the empirical data. The path loss exponent obtained for the LOS scenarios is found to be very close to that of the free space path loss model.\looseness=-1
\end{abstract}

\vspace{-1mm}
\section{Introduction}
\blfootnote{This work has been supported in part by NASA under the Federal Award ID number NNX17AJ94A and by DOCOMO Innovations, Inc.}

The requirement for the high data rates in fifth generation cellular systems can be met using large bandwidths on the order of GHz~\cite{Shafi2018}. Since the sub-6~GHz spectrum is currently overutilized, higher-frequency bands, such as millimeter-wave~(mmWave) bands, have attracted growing attention among the researchers. Deploying wireless systems requires an accurate understanding of the channel propagation characteristics in the deployment band. Statistical channel models for sub-6~GHz cellular systems are derived based on extensive channel measurement campaigns. However, characteristics of the wireless channel are different at different frequencies. Therefore, similar efforts should also be put in place to characterize the nature of radio propagation at mmWave frequencies~\cite{khatun&globalsip:2018}.

One of the main drawbacks of the mmWave wireless channel is the high path loss. To address this problem, high gain antennas are used, and the antennas are aligned based on the the strongest path to further improve the signal strength. The channel sounders used for modeling the sub-6~GHz channel mainly measure the power delay profile (PDP) of the channel. However, due to the high path loss in the mmWave bands, angular profile of the channel, known as the power angular-delay profile (PADP), should also be measured~\cite{Lin2017}. \looseness =-1

Several researchers and companies have been carrying out channel measurements at mmWave frequencies in different environments, e,g., in a laboratory environment~\cite{7228884}, a five-story building~\cite{7954664}, and an office environment~\cite{7880929}. However, there is still a lack of measurement results in large and crowded indoor environments. 

In this paper, we use a rotating directional horn antennas (RDA)-based channel sounder to characterize the mmWave channel~\cite{wahab_outdoor} at 28~GHz. We performed measurement campaigns in a library environment considering various line-of-sight~(LOS) and non-line-of-sight~(NLOS) scenarios. We present the PADP model and two path loss models, namely, the floating-intercept model~(FIM) and the close-in free space reference distance model~(CIM). We obtain the parameters of the path loss models based on the empirical data. Results show that both path loss models provide good estimates, and the path loss exponent~(PLE) determined from the CIM is very close to free space PLE. 

The rest of the paper is organized as follows. In Section~\ref{Sec:Measurement}, we describe the channel sounder hardware and the measurement environment. In Section~\ref{Sec:Modeling}, we present the PADP model and the path loss models. Measurement results are provided in Section~\ref{Sec:Results}, and the paper is concluded in Section~\ref{sec:Conc}.

\begin{figure}[t!]
\centering
\centerline{\includegraphics[trim=.5cm 9cm 8.5cm 0.35cm, clip,width=\linewidth]{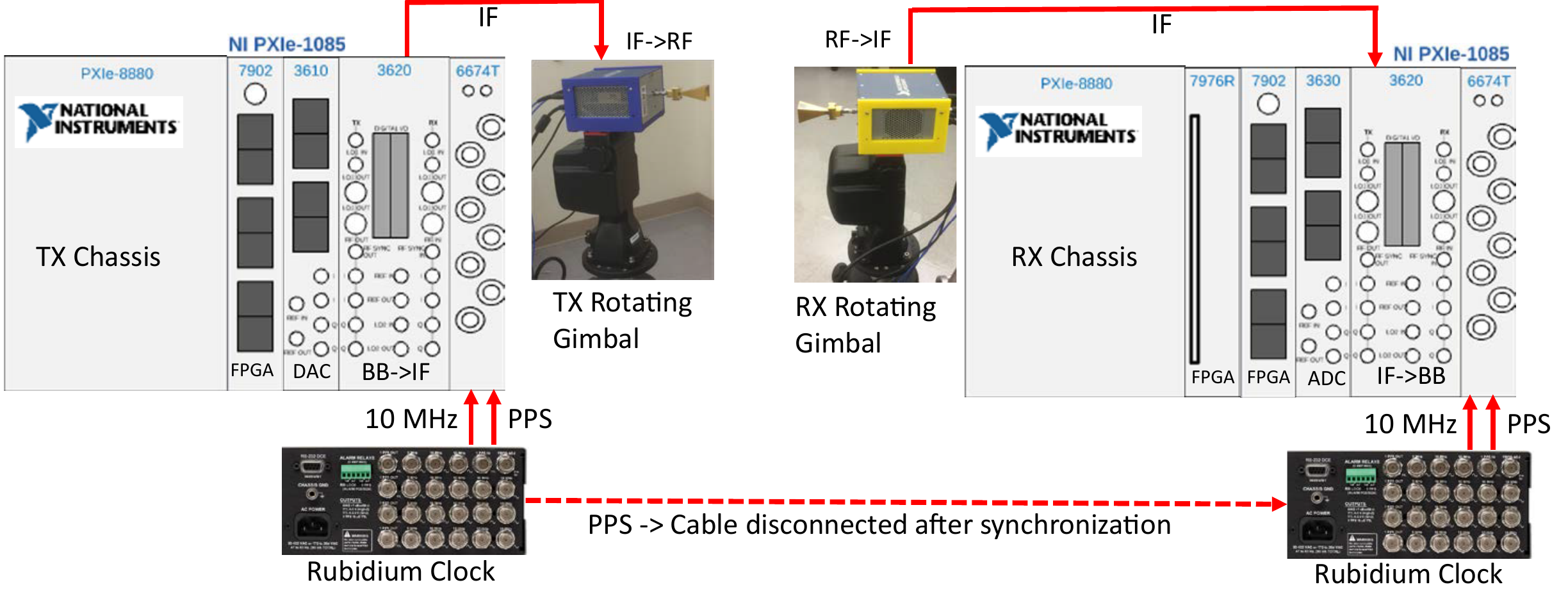}}
\caption{28~GHz channel sounder TX and RX hardware setup with two Rb clocks.}\label{Fig:pxisetup}
\vspace{-5mm}
\end{figure}

\section{Measurement Setup and Environment}
\label{Sec:Measurement}
% In this section, we describe the measurement setup and library environment. 
\subsection{Measurement Hardware}
\label{Sec:Hardware}
The measurements are performed using an NI-based channel sounder~\cite{NImmwave} shown in Fig.~\ref{Fig:pxisetup}. The sounder hardware consists of NI PXIe-1085 TX/RX chassis, 28~GHz TX/RX mmWave radio heads from NI, FS725~Rubidium~(Rb) clocks~\cite{SRS}, and FLIR~PTU-D48E gimbals~\cite{FlirSystems}. The 10~MHz and pulse per second~(PPS) signals generated by the Rb clocks are connected to PXIe~6674T modules at the TX and the RX. Directional horn antennas~\cite{sageM}, with 17~dBi gain, and $26^{\circ}$/$24^{\circ}$ beam-width in the elevation/azimuth plane, are connected to the mmWave radio heads. The radio heads are placed on the rotating gimbals to be able to measure the PADP of the channel. Three different elevation angles $\{-20^\circ, 0^\circ, 20^\circ\}$ and 19 different azimuth angles ranging from $-167.98^\circ$ to $167.98^\circ$, with increments of $20^\circ$, are scanned at all measurements. The TX power is set to $P_\mathrm{TX}=-10$~dBm.\looseness =-1

The LabVIEW-based sounder code periodically transmits a Zadoff-Chu (ZC) sequence of length 2048 to sound the channel. The ZC sequence is over-sampled by 2 when the sounder operates on 2 GHz bandwidth. Then, it is filtered by the root-raised-cosine filter, and the generated samples are uploaded to PXIe-7902 FPGA. These samples are sent to PXIe-3610 digital-to-analog converter with a sampling rate of $f_s=3.072$~GS/s. The PXIe-3620 module up-converts the base-band signal to IF, and the 28~GHz mmWave radio head up-converts the IF signal to RF. This process is reversed at the RX side, and the complex CIR samples are sent to the PXIe-8880 host PC for further processing. The channel sounder provides 0.651~ns resolution in the delay domain. The dynamic range of the analog-to-digital converter is $60$~dB, and the maximum measurable path loss is $185$~dB.
\vspace{-0.2mm}
\subsection{Measurement Environment}
\label{Sec:Environment}
\begin{figure}[!t]
	\centering
% 	\begin{subfigure}{\columnwidth}
% 	\centering
% 	\centerline{\includegraphics[trim=0cm 0cm 0cm 0cm, clip,width=0.95\linewidth]{Library_Satellite.JPG}}
% 	\caption{}
%     \end{subfigure}	

	\begin{subfigure}{\columnwidth}
	\centering
	\centerline{
	\includegraphics[trim=0cm 0cm 0cm 0cm, clip,width=0.85\linewidth]{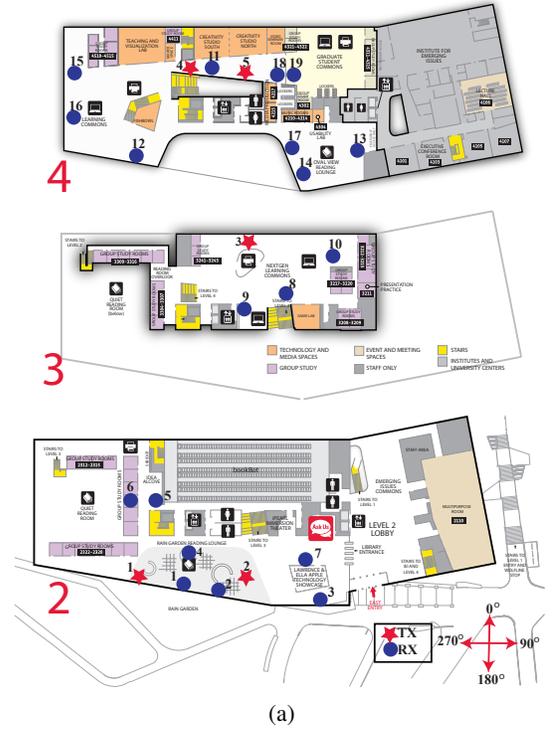}
	}
	\caption{}
    \end{subfigure}
    
    \begin{subfigure}{\columnwidth}
	\centering
	\centerline{\includegraphics[trim=3cm 3cm 1.5cm 3cm, clip,width=0.85\linewidth]{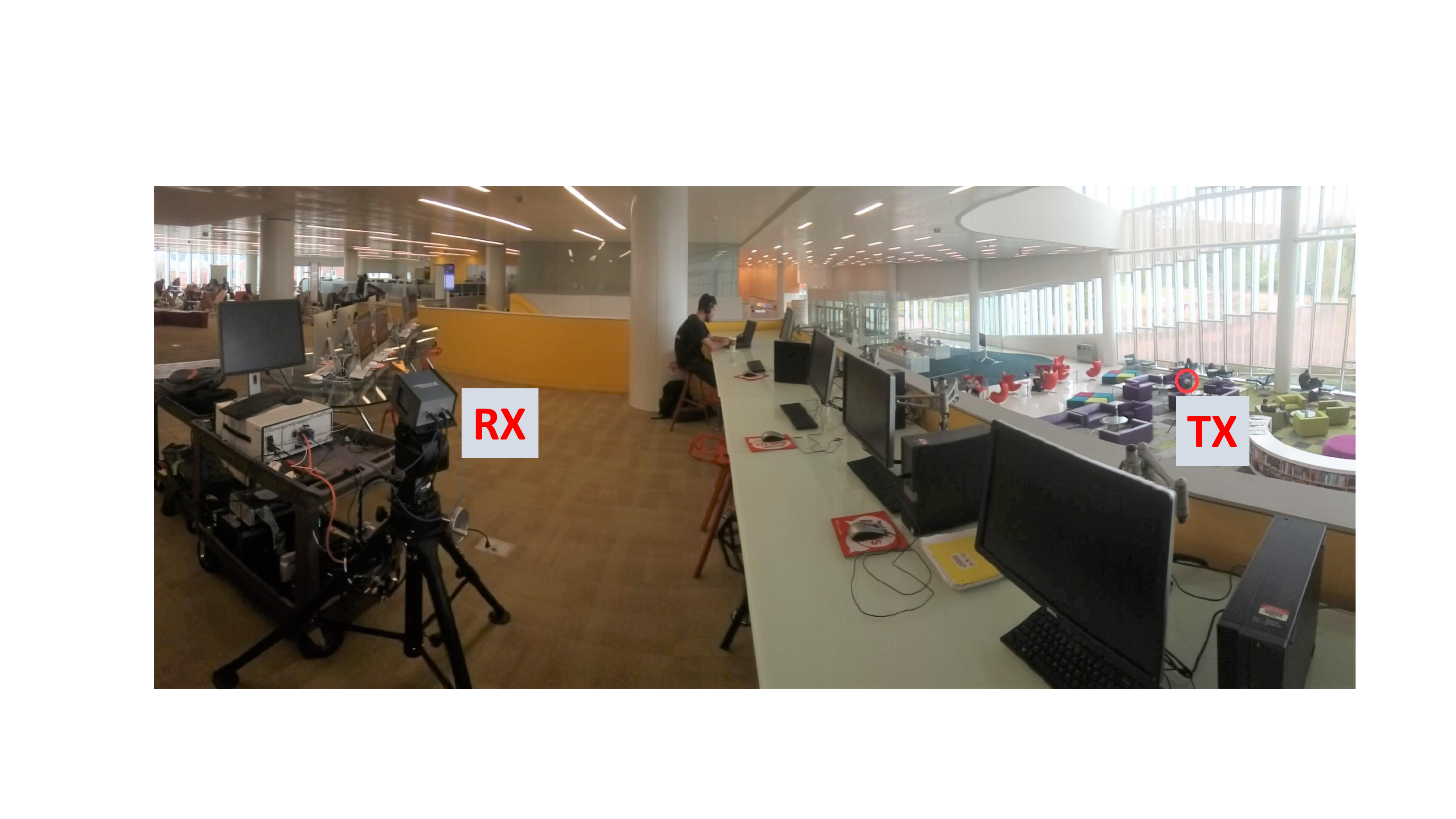}}
	\caption{}
    \end{subfigure}
    
     \caption{
     The Hunt library 
    %  floor plans and the TX/RX locations.
    (a) floor plans with TX/RX locations and (b) snapshot of the measurement scenario for TX1-RX3 pair.
     }
     \label{Fig:Meas_Env}
     \vspace{-5.5mm}
\end{figure}

The indoor measurements are conducted at three different floors at the Hunt Library of NCSU, which is about 140~m in length, 55~m at its widest point, and 27~m high at tallest point. The library is furnished with chairs, desks, and tables of different materials, and movable whiteboards. In addition, the desks are equipped with computers and electronic devices. The walls and floors are made of concrete, and the group study rooms have glass doors. The layout of the measurement environment with the TX/RX locations is shown in Fig.~\ref{Fig:Meas_Env}(a). Nineteen indoor measurements, including scenarios where the TX and RX are placed on different floors, are performed for five different TX locations and TX-RX separation distances ranging from 10~m to 50~m for both LOS and NLOS scenarios (see Fig.~\ref{Fig:Meas_Env}(b) for a sample measurement scenario). In four of the scenarios, the link is obstructed by glass, and these scenarios are also considered as NLOS. The TX and the RX were fixed at a height of 1.8~m and 1.5~m from the ground, respectively. 
\vspace{-0.2mm}
\section{PADP and Path Loss Modeling}
\label{Sec:Modeling}
\subsection{PADP Model}
\label{Sec:PADP}
For measurements at TX-RX separation distances greater than 30~m, the synchronization cable between the clocks needs to be disconnected, causing the problem known as the clock drift. In addition, the RDA-based channel sounders suffer from another problem caused by the antenna rotations. As the horn antennas are rotated during a measurement, the total distance travelled by a multipath component~(MPC) changes, which may result in detecting ghost MPCs or missing some actual MPCs. These effects in the delays of the peaks detected from the PDPs are corrected using the algorithm described in~\cite{erden2019}. Next, the MPCs are extracted using the peak search algorithm~\cite{erden2019}. The PADP of the channel, where the AoAs and the AoDs of the MPCs over all possible TX/RX angles are estimated as well, can be expressed as: \looseness=-1
\begin{align}
\label{PADP:eq}
PADP(\tau, \bm{\theta}^{\mathrm{AoD}} , \bm{\theta}^{\mathrm{AoA}}) = &\sum_{n=1}^{N}  \alpha_n \delta (\bm{\theta}^{\mathrm{AoD}}-\bm{\theta}^{\mathrm{AoD}}_n)  \\ \nonumber &\times  \delta (\bm{\theta}^{\mathrm{AoA}}-\bm{\theta}^{\mathrm{AoA}}_n) \delta(\tau-\tau_{n}),
\end{align}
where $\bm{\theta}^{\mathrm{AoD}}_n = [\theta^{\mathrm{AoD,Az}}_n \,\, \theta^{\mathrm{AoD,El}}_n]^{\sT}$ is the two-dimensional AoD of the $n$th MPC at the TX in the azimuth and elevation planes, $\bm{\theta}^{\mathrm{AoA}}_n = [\theta^{\mathrm{AoA,Az}}_n \,\, \theta^{\mathrm{AoA,El}}_n]^{\sT}$ is the two-dimensional AoA of the same MPC at the RX, $\alpha_n$ is the path gain, $\tau_n$ is the delay of the $n$th MPC, and $N$ is the total number of MPCs considering all possible TX/RX directions.\looseness =-1

\subsection{Path Loss Models}
\label{Sec:Pathloss}
To calculate the path loss from the measured data, first, the omnidirectional received power is calculated by summing the linear power over all angles, i.e.,
\begin{equation}
    P_{\mathrm{RX}}(d)=\sum_{i}\sum_{j,k}PADP(\tau_i, {\theta}_j^{\mathrm{AoD}} , {\theta}_k^{\mathrm{AoA}}),
\end{equation}
where $i,j$, and $k$ are the indices of the delay and angles of the corresponding PADP. Then, the path loss can be calculated by
\begin{equation}
    PL(d)[\mathrm{dB}]=P_{\mathrm{TX}}-10\log_{10}(P_{\mathrm{RX}}(d))+G_{\mathrm{TX}}+G_{\mathrm{RX}},
\end{equation}
where $G_{\mathrm{TX}}$ and $G_{\mathrm{RX}}$ are TX and RX antenna gains in dBi. 
Given the measured data, parameters of the two path loss models can be calculated. The CIM has the following form~\cite{7109864}:
\begin{align}
\label{eq:CIM}
\nonumber \mathrm{PL}^\mathrm{CIM}(f,d)[\mathrm{dB}]=&\mathrm{FSPL}(f,d_0)+10n\log_{10}\left(\frac{d}{d_0}\right)
\\
&+\chi^\mathrm{CIM}_\sigma, \quad \textrm{for } d\ge d_0, 
\end{align}
where $d_0$ is the close-in free space reference distance and set to 1~m in this study, $\mathrm{FSPL}(f,d_0)$ is the free space path loss at distance $d$, $n$ is the PLE, and $\chi^\mathrm{CIM}_\sigma$ is a lognormal random variable with 0~dB mean and $\sigma$ standard deviation. 
The FIM is given as~\cite{7109864}
\begin{align}
\label{eq:FIM}
\mathrm{PL}^\mathrm{FIM}(d)[\mathrm{dB}]&=\alpha + 10 \beta \log_{10}(d)+\chi^\mathrm{FIM}_\sigma,
\end{align}
where $\alpha$ is the intercept in~dB and $\beta$ is the slope, and $\chi^\mathrm{FIM}_\sigma$ is a lognormal random variable with 0~dB mean and $\sigma$ standard deviation. 
\subsection{Delay Spread}
It is common in the literature to characterize the delay dispersion using the metric known as the root-mean-square delay spread (RMS-DS)~\cite{7857002}. The RMS-DS is calculated based on the extracted MPCs as follows 
% The PDP can be obtained as follows
% \begin{equation}
%     PDP(\tau)=\max_{\bm{\theta}^{\mathrm{AoD}}}\max_{\bm{\theta}^{\mathrm{AoA}}}PADP(\tau, \bm{\theta}^{\mathrm{AoA}} , \bm{\theta}^{\mathrm{AoA}}).
% \end{equation}
% Then, the RMS-DS is
% \begin{equation}
%     \tau_{\mathrm{rms}}={\sqrt{\frac{\sum_{n=1}^NPDP(\tau_n)(\tau_n-\tau_{avg})^2}{\sum_{n=1}^N PDP(\tau_n)}}},
% \end{equation}
\begin{equation}
    \tau_{\mathrm{rms}}={\sqrt{\frac{\sum_{n=1}^N P_n(\tau_n-\tau_{avg})^2}{\sum_{n=1}^N P_n}}},
\end{equation}
where $P_n$ is the power of the $n$th MPC, and $\tau_{avg}$ is the mean delay given by
\begin{equation}
    \tau_{\mathrm{avg}}=\frac{\sum_{n=1}^N P_n\tau_n}{\sum_{n=1}^N P_n}.
\end{equation}
% \begin{equation}
%     \tau_{\mathrm{avg}}=\frac{\sum_{n=1}^N PDP(\tau_n)\tau_n}{\sum_{n=1}^N PDP(\tau_n)}.
% \end{equation}

\begin{figure}[t]
	\centering
	\begin{subfigure}{6.5cm}
	\centerline{\includegraphics[trim=0cm 0cm 0cm 0cm, clip,width=\linewidth]{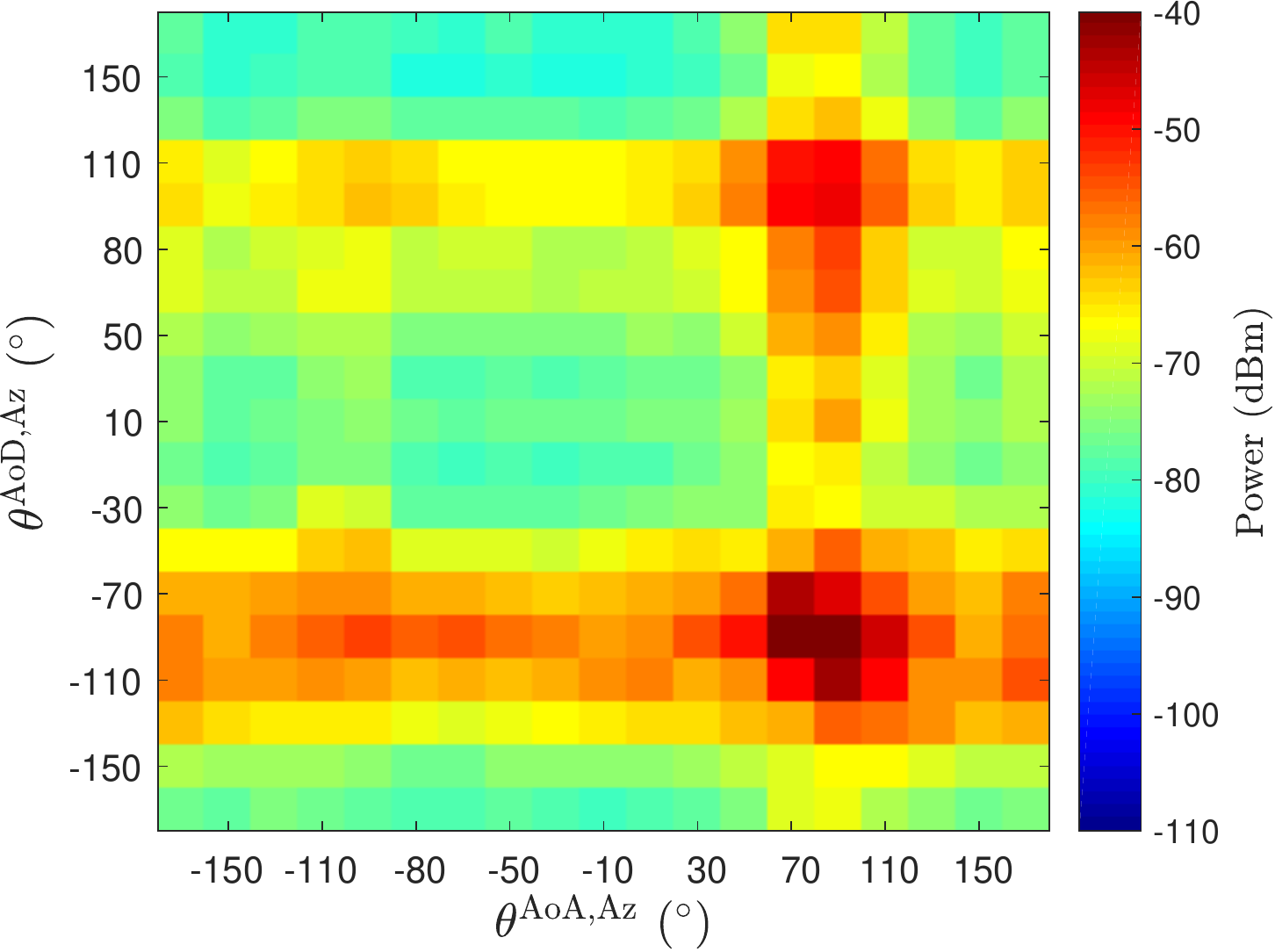}}
	\caption{}
    \end{subfigure}	
    % \quad
	\begin{subfigure}{6.5cm}
	\centering
	\centerline{\includegraphics[trim=0cm 0cm 0cm 0cm, clip,width=\linewidth]{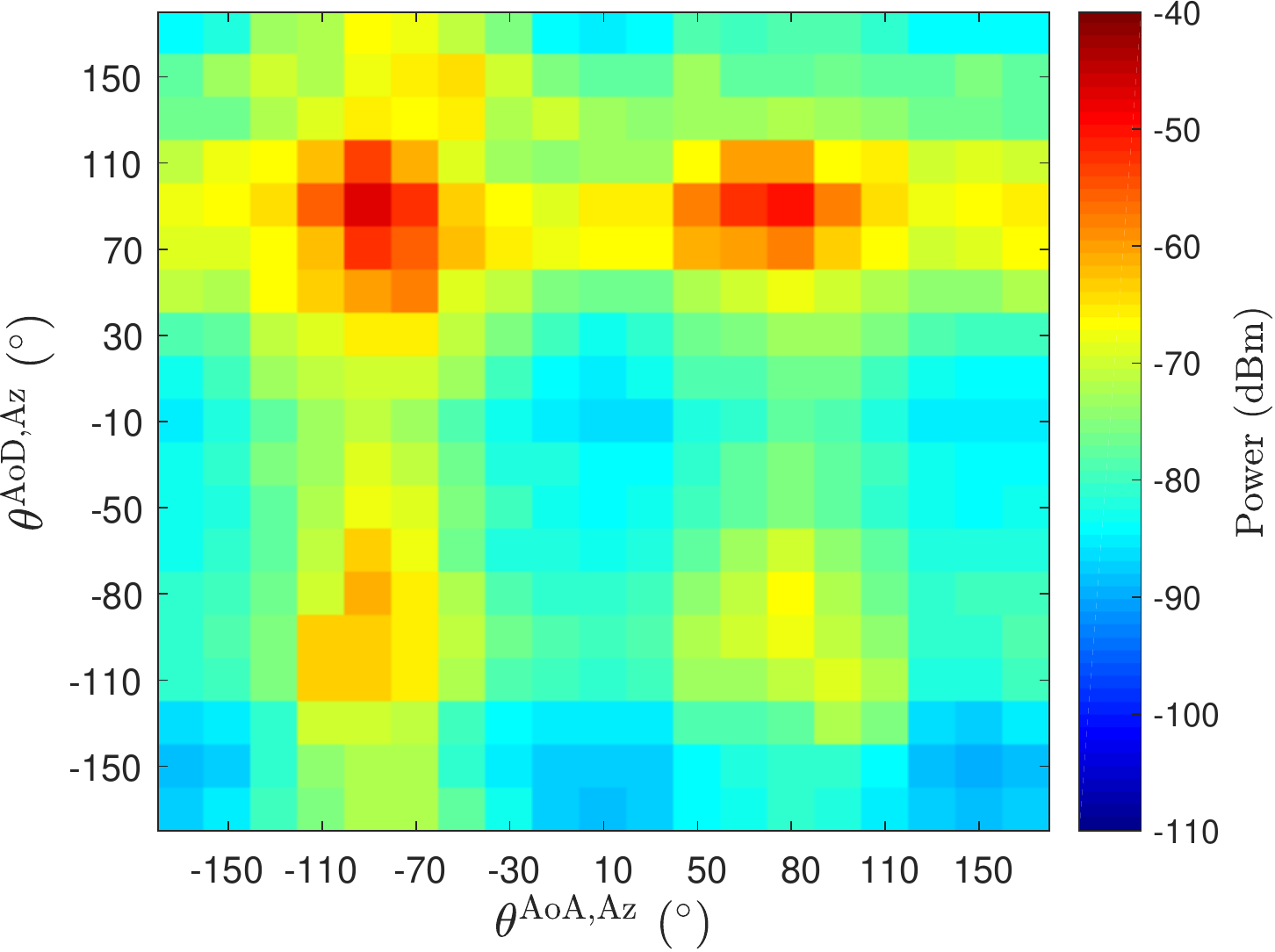}}
	\caption{}
    \end{subfigure}	
    
	\begin{subfigure}{6.5cm}
	\centering
	\centerline{\includegraphics[trim=0cm 0cm 0cm 0cm, clip,width=\linewidth]{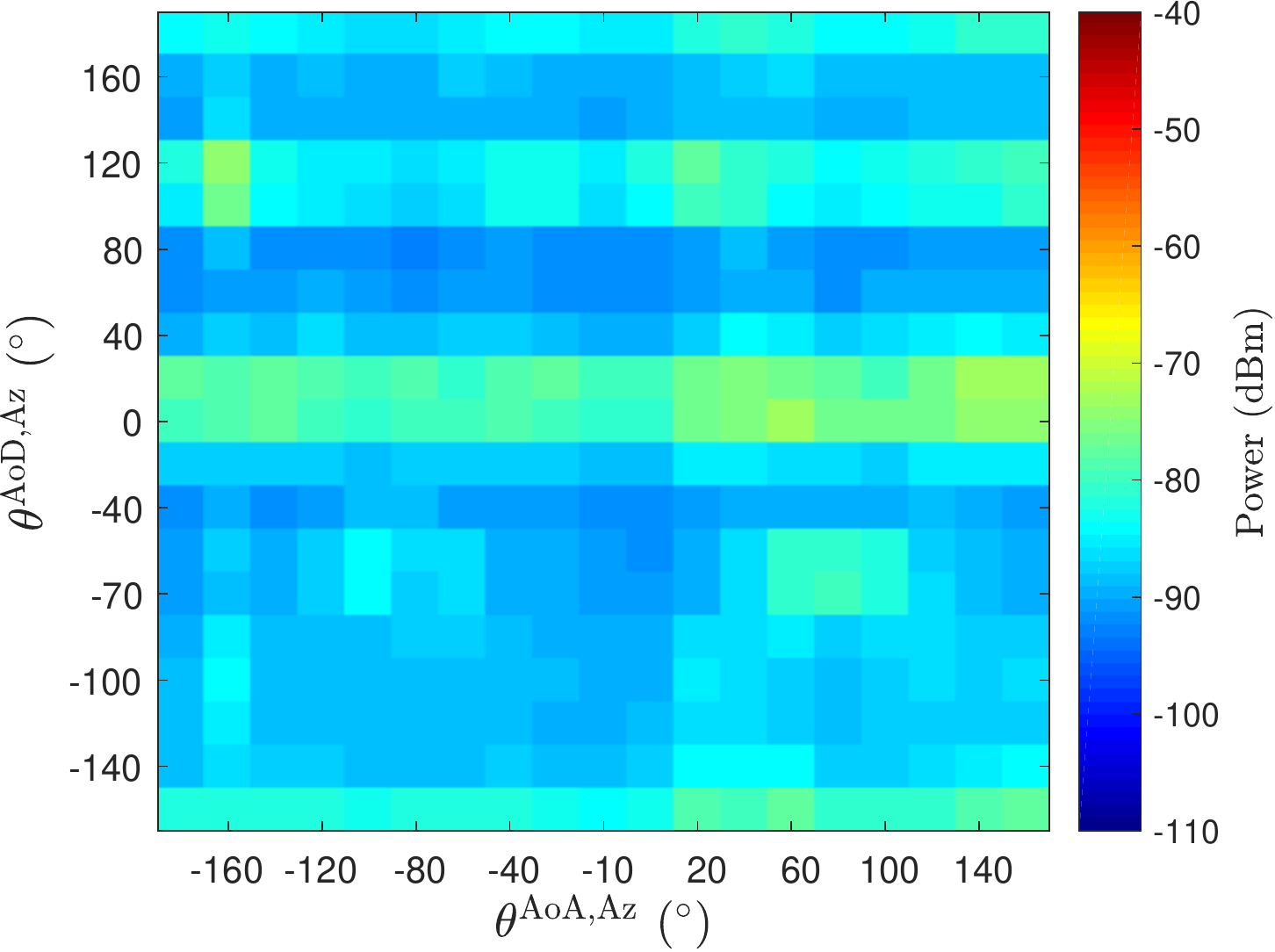}}
	\caption{}
    \end{subfigure}
    
     \caption{PADPs for (a) LOS: TX4-RX15, (b) NLOS-glass: TX5-RX19, and (c) NLOS: TX1-RX6 scenarios. TX $El = 0^\circ$, RX $El = 0^\circ$.}
     \label{Fig:PADP}
     \vspace{-5mm}
\end{figure}

\section{Measurement Results}
\label{Sec:Results}
\subsection{PADP Measurements}
The PADPs extracted from three sample scenarios, namely, LOS, NLOS-glass, and NLOS, are shown in Fig.~\ref{Fig:PADP}(a), Fig.~\ref{Fig:PADP}(b), and Fig.~\ref{Fig:PADP}(c), respectively, for fixed TX/RX elevation angles. The NLOS scenario refers to the case where the link is obstructed by walls and glass doors, whereas the NLOS-glass scenario represents the case where the obstruction is only due to a glass door. TX-RX separation distance is about 21~m in each case. The LOS received power is greater than the NLOS and NLOS-glass cases, as expected. A significant difference in the received power is observed between the NLOS and NLOS-glass scenarios, indicating that the walls introduce a significantly higher attenuation when compared with the glass doors.\looseness =-1

It is also observed that the received power is mainly concentrated in two departure angles for the LOS and NLOS-glass scenarios, corresponding to the LOS path and first-order reflections. On the other hand, for the NLOS scenario, the received power is spread more evenly over many directions corresponding to the first and higher-order reflections. 

% \begin{figure}[t]
% 	\centering
% 	\begin{subfigure}{4cm}
% 	\centerline{\includegraphics[trim=0cm 0cm 0cm 0cm, clip,width=\linewidth]{TX4RX15_Power_2D_v2.eps}}
% 	\caption{}
%     \end{subfigure}	
%     % \quad
% 	\begin{subfigure}{4cm}
% 	\centering
% 	\centerline{\includegraphics[trim=0cm 0cm 0cm 0cm, clip,width=\linewidth]{TX5RX19_Power_2D_v2.eps}}
% 	\caption{}
%     \end{subfigure}	
    
% 	\begin{subfigure}{4cm}
% 	\centering
% 	\centerline{\includegraphics[trim=0cm 0cm 0cm 0cm, clip,width=\linewidth]{TX1RX6_Power_2D_v2.eps}}
% 	\caption{}
%     \end{subfigure}
    
%      \caption{PADPs for (a) LOS: TX4-RX15, (b) NLOS-glass: TX5-RX19, and (c) NLOS: TX1-RX6 scenarios. TX $El = 0^\circ$, RX $El = 0^\circ$.}
%      \label{Fig:PADP}
%      \vspace{-3mm}
% \end{figure}

\subsection{Path Loss Results}
\begin{figure}[t]
	\centering
	\begin{subfigure}{\columnwidth}
	\centering
	\centerline{\includegraphics[trim=0cm 0cm 0cm 0cm, clip,width=0.75\linewidth]{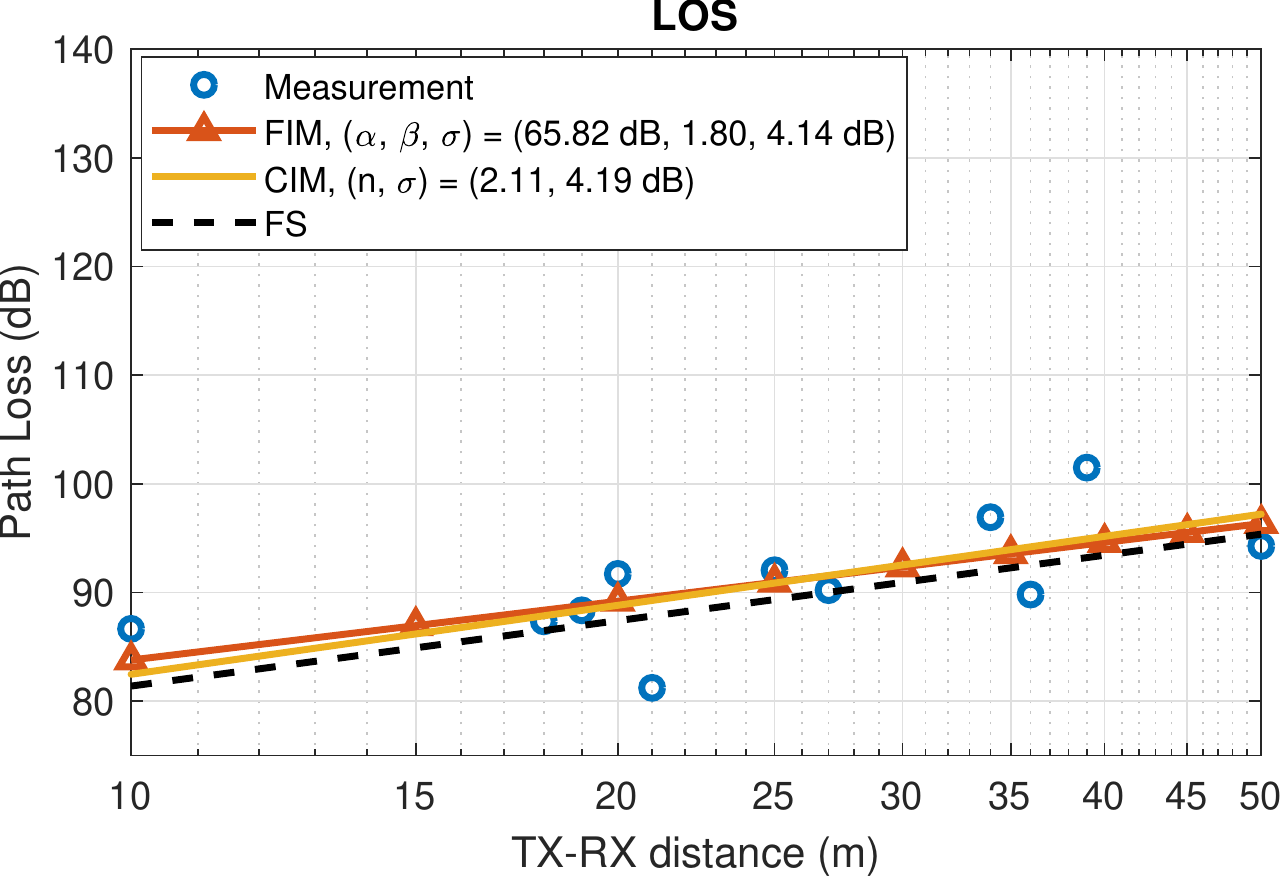}}
	\caption{}
    \end{subfigure}	

	\begin{subfigure}{\columnwidth}
	\centering
	\centerline{\includegraphics[trim=0cm 0cm 0cm 0cm, clip,width=0.75\linewidth]{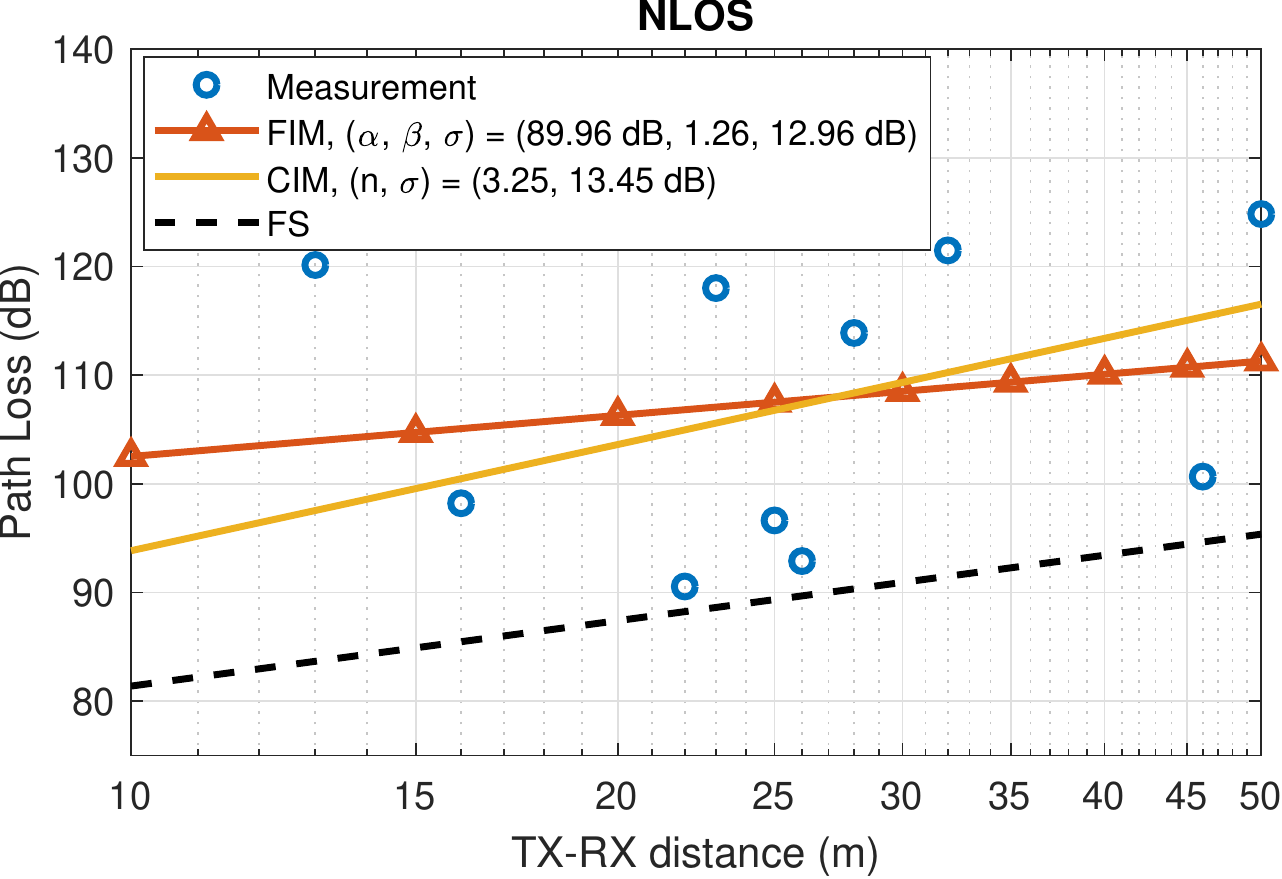}}
	\caption{}
    \end{subfigure}
    
     \caption{Path loss results from the measurement data, and FIM and CIM models for (a) LOS and (b) NLOS scenarios.}
     \label{Fig:Pathloss}
     \vspace{-3mm}
\end{figure}

The empirical path loss results and the corresponding fitted models are shown in Fig.~\ref{Fig:Pathloss}(a) and Fig.~\ref{Fig:Pathloss}(b) for LOS and NLOS scenarios, respectively. It turns out that, based on the shadowing factors~($\sigma$), the two models have similar performance in fitting the measured data. The LOS PLE determined by the CIM is 2.11, which is very close to the free space PLE, $n=2$. The relatively higher deviations of some of the empirical data from the free space propagation can be attributed to the waveguide effect caused by the hallways or the human blockages. In NLOS, PLE is found to be 3.25, indicating higher signal level degradation. The higher $\sigma$ in NLOS can be attributed to differences in channel types (e.g., material types).

% As it is common in the literature, we also investigate the root-mean-square delay spread (RMD-DS).

% \begin{table}[t]
% % \footnotesize
% % \renewcommand{\arraystretch}{1.2}
% \caption {Pros/Cons of Different Channel Sounder Types. }
% \label{Tab:sounders}
% \resizebox{0.48\textwidth}{!}
% {\begin{tabular}{c|c|c}
% \hline
%  Sounder type & Pros &  Cons \\\hline
% Gimbal~\cite{MacCartney,Lin2017} & simple, low cost & slow  \\
% Switching antenna~\cite{Papazian2016} & fast & expensive, bulky  \\
% Phased array~\cite{Bas2018} & fast &  expensive, needs calibration \\\hline
% % \vspace{-3mm}
% \end{tabular}}
% \end{table}

% \begin{figure}[t!]
% \centering
% \centerline{\includegraphics[trim=0cm 0cm 0cm 0cm, clip,width=0.9\linewidth]{RMS_DS}}
% \caption{CDF of the RMS-DS for LOS and NLOS links.}\label{Fig:RMS-DS}
% \end{figure}
\vspace{-0.29mm}
\subsection{Delay Spread Statistics}
\begin{figure}[t!]
\centering
\centerline{\includegraphics[trim=0cm 0cm 0cm 0cm, clip,width=0.75\linewidth]{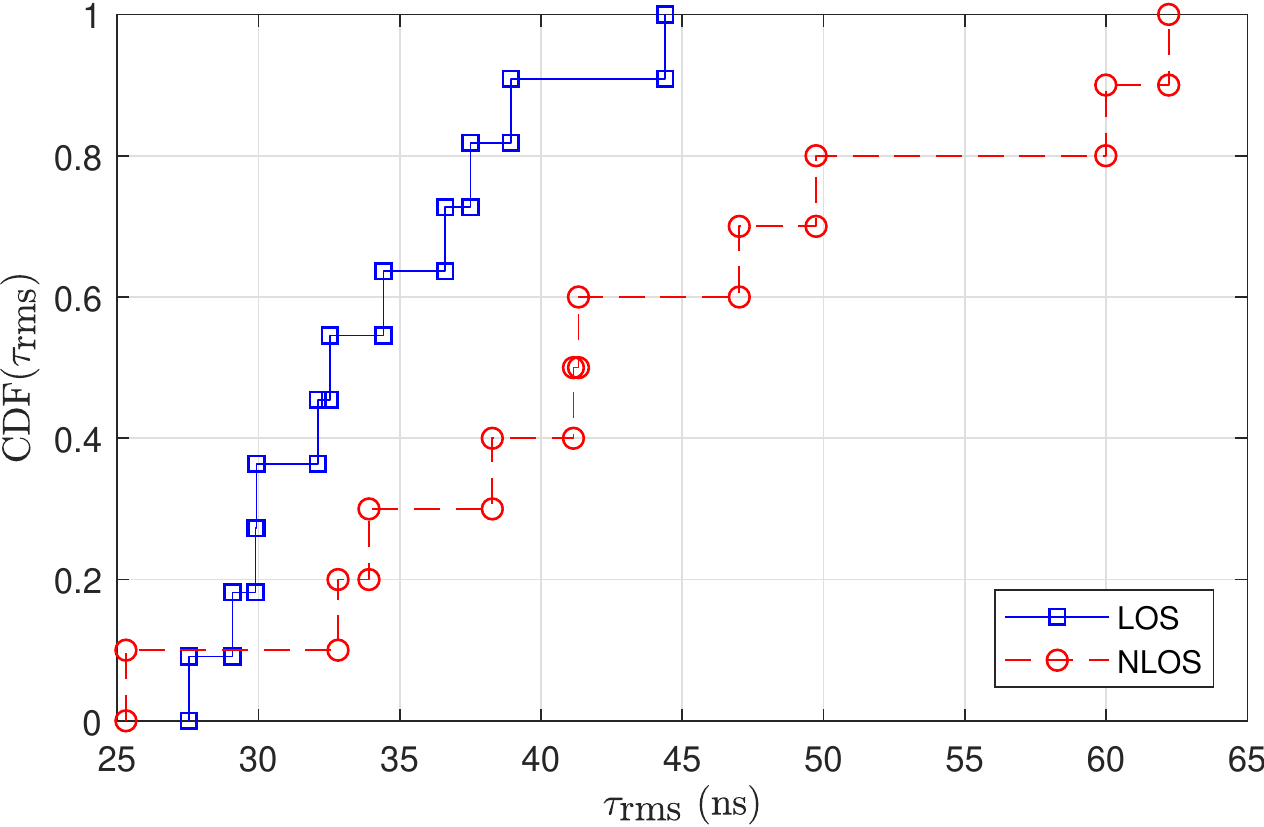}}
\caption{CDFs of the RMS-DS for the LOS and NLOS links.}\label{Fig:RMS_DS}
\vspace{-5mm}
\end{figure}

The cumulative distribution functions (cdfs) of the RMS-DS for the LOS and NLOS links are shown in Fig.~\ref{Fig:RMS_DS}. The mean RMS-DS is calculated to be 33.9~ns and 43.1~ns for the LOS and NLOS links, respectively. The mean RMS-DS in the NLOS case is significantly greater than that of the LOS case because the LOS path, which is the strongest path and arrives the earliest, is blocked in NLOS scenarios, shifting the mean of the delay dispersion to the right.\looseness=-1

\section{Conclusions}
In this paper, mmWave channel measurements at 28~GHz are presented for a library environment. The library is selected to investigate the channel characteristics in large and open indoor environments that are built and furnished with different types of materials. It is shown that both the FIM and the CIM provide a good match with the empirical path loss results. It is observed that when the link is obstructed by a glass door, the power of the LOS path decreases notably (7~dB difference was noted between the measurements with TX5-RX18 and TX5-RX19), and when the link is obstructed by a concrete wall, the attenuation is more severe.
\label{sec:Conc}
\section*{Acknowledgment}
The authors would like to thank the graduate students Harini Narasimhan, Uddeshay Gupta, Martins Ezuma, Vageesh A. Dambal, Nishant Anand, Chunmay Datar, Sri Latha Sunkara, Madhusudhan Gopalappa, and Kairui Du for their help with the measurements. 
% \balance
\bibliography{IEEEabrv,references}
\bibliographystyle{IEEEtran}
\end{document}